\documentclass[11pt]{article}
\usepackage{amssymb,amsmath,amsfonts,amsthm,lscape}
\usepackage{booktabs}

\usepackage{enumitem}
\usepackage{color}
\theoremstyle{plain}

\theoremstyle{definition}

\begin{document}

\centerline{\Large \bf Miura-reciprocal transformations} \vskip 0.25cm
\centerline{ \Large \bf for non-isospectral Camassa-Holm hierarchies}\vskip 0.25cm
\centerline{ \Large \bf in $2+1$ dimensions}

\vskip 0.5cm
\centerline{P.G. Est\'evez and C. Sard\'on}
\vskip 0.5cm
\centerline{Department of Fundamental Physics, University of
Salamanca,}
\centerline{Plza. de la Merced s/n, 37.008, Salamanca, Spain.}

\vskip 1cm

\begin{abstract}
We present two hierarchies of partial differential equations in $2+1$ dimensions. Since there exist reciprocal transformations
that connect these hierarchies  to the Calogero-Bogoyavlenski-Schiff equation and its modified version, we can prove that one of the  hierarchies can be considered as a modified version of the other. The connection between them can be achieved by means of a combination of reciprocal and Miura transformations.
\end{abstract}

\section{Introduction}
Is there any way to decide whether an integrable nonlinear equation is actually new, or merely a
disguised form of another member of the class of the integrable zoo? The question is far from being trivial,
given the amount of papers published on the subject and the number of \textit{different}  names often given to the
\textit{same} equation in the current literature. A clear procedure has to be found in order to relate the
allegedly new equation to the rest of its integrable siblings. This letter is an attempt to shed some
light on the solution of this seemingly simple, but nonetheless crucial, question.

 According to the Painlev\'e criteria \cite{painleve}, a nonlinear PDE is said
to be integrable when its solutions turn out to be single valued in a neighborhood of the movable singularities
manifold. As is well known, the existence of this property can be checked through an algorithmic procedure which is
valid both in the case of nonlinear ordinary or partial differential equations \cite{wtc}. For this so
called Painlev\'e property, a method has been developed by Weiss \cite{weiss83} that allows us to determine
the set of solutions that give rise to a truncated Painlev\'e expansion. In this situation, one can show that
the manifold of movable singularities should satisfy a set of equations named \textbf{singular manifold
equations} (SME). Long experience based on many already worked examples by the present and other authors
\cite{estevez23}, \cite{estevez25}, \cite{estevez04} indicates that SME
could be the canonical form that identifies an integrable partial differential equation (PDE). If this were the case, two seemingly different integrable PDEs with the same set of SME must be related through some
transformation. The remaining point is the nature and identification of such a transformation.

There is a caveat in this project that we shall  deal with  first. The standard form of the Painlev\'e test
cannot be applied to some interesting PDEs. Therefore one cannot be sure of whether in those cases the equation is
or is not single-valued in the initial conditions. This is the case \cite{gp95}, for instance, of the
celebrated Camassa-Holm equation \cite{camassa}. This equation has been known to be integrable for some time and
it has an associated linear problem. Therefore, discarding any kind of pathology, one should be able to write
down a transformation that brings it into a form in which the Painlev\'e test can be applied. This is the case of the relation between the Harry-Dym and Korteweg de Vries equation that appears in \cite{ibragimov}. Reciprocal transformations for peakons equations like Camassa-Holm and Degasperis-Procesi were derived in \cite{hone} to transform these equations to equations with the Painlev\'e property. Actually, reciprocal transformations have proved to be a powerful tool to relate different equations and conservation laws  \cite{euler1}, \cite{euler2}, \cite{rogers2}, \cite{rogers3}. In chapter 3 of reference \cite{rogers1}, reciprocal relations for gasdynamics equations as transformations of B\"acklund type were derived.

For equations in $2+1$ dimensions, it was proved in \cite{rogers5} that there exists a reciprocal link between the  Harry-Dym equation in $2+1$ dimensions and the singular manifold equations \cite{weiss83} of the Kadomtsev-Petviashvili equation. Reciprocal transformations for $2+1$ shallow water equations were identified in \cite{h00}. In reference \cite{estevez05-1} we introduced an
integrable generalization to $2+1$ dimensions of the Camassa-Holm hierarchy (CH2+1). By using reciprocal
transformations, we were able to show explicitly that the n-component of the hierarchy was in fact equivalent
to n copies of the Calogero-Bogoyavlenski-Schiff (CBS) equation \cite{cbs}, \cite{pick}. This CBS equation
possesses the Painlev\'e property and the singular manifold method can be applied to obtain its Lax pair and
other relevant properties \cite{estevez05-1}.

It is tempting to assume that these and similar results appearing in other recent
publications \cite{estevez09} are all but coincidence. The reciprocal transformations should be a useful
method to transform equations in which the Painlev\'e test cannot be applied to other fully tractable equations  through the
singular manifold method (SMM).

To convince the reader that this conjecture has some grounds, we have  recently presented another non-trivial
example of the procedure. Qiao has reported an integrable equation \cite{qiao} for which the Painlev\'e test
is neither applicable nor constructive. This equation is the second member of a hierarchy in $1+1$ dimensions
\cite{qiao2007} that one of us has  recently generalized to $2+1$ dimensions \cite{estevez51}. One can prove that a reciprocal transformation exists which, as in the CH2+1 case, allows us
to transform the n-component of the hierarchy into n copies of the modified Calogero-Bogoyavlenskii-Schiff
(mCBS), which is known to have the Painlev\'e property \cite{estevez04}. The last step can be shown by
transforming mCBS into CBS by means of a Miura transformation. The singular manifold equations are the same for
both sets of equations. We shall denote in the following the hierarchy introduced in \cite{estevez51} as mCH(2+1) because it can be considered as a modified version of the CH(2+1) hierarchy introduced in \cite{estevez05-1}.

To summarize: CH(2+1) and mCH(2+1) are directly related to CBS and mCBS respectively through two different sets
of reciprocal transformations.
Aside from this property, there exists a Miura transformation relating CBS and mCBS. The relationship between these two hierarchies
must necessarily include both Miura and reciprocal transformations. Therefore, for obvious reasons, the name Miura-reciprocal
transformations will be  used extensively throughout the paper and they will be the subject of it.

\section{Reciprocal transformations for hierarchies}
 In this section we shall  briefly summarize and improve the results of \cite{estevez05-1} and
\cite{estevez51} in order to establish the reciprocal transformation that connects the CH(2+1) and mCH(2+1)
hierarchies with CBS and mCBS respectively. Many details (especially those referring to the detailed
calculation) are omitted and can be obtained in the above cited references.

It could be useful from the beginning to say that we shall use capital letters for the dependent and
independent variables connected with CH(2+1). Henceforth, lower case letters  will be used  for mCH(2+1).

\subsection{CH(2+1) versus CBS}
The CH(2+1) hierarchy can be written in a compact form as:
\begin{equation}
U_T=R^{-n}U_Y, \label{1}
\end{equation}
where $R$ is the recursion operator defined as:
\begin{equation}
R=JK^{-1},\quad K=\partial_{XXX}-\partial_X,\quad J=-\frac{1}{2}\left(\partial_XU+U\partial_X\right),\quad \textrm{where} \quad \partial_X=\frac{\partial}{\partial X}.
\label{2}
\end{equation}
Note that the factor $-\frac{1}{2}$ in the definition of $J$ is not essential and it has been introduced to
make the later identification between the time variables easier.

This hierarchy was introduced in \cite{estevez05-1} as a generalization of the Camassa-Holm hierarchy. The recursion operator is the same as in the $1+1$ dimensional Camassa-Holm hierarchy. From this point of view, the spectral problem is the same \cite{calogero} and the $Y$-variable is just another ``time" variable  \cite{h00}, \cite{ivanov}.

The $n$ component of this hierarchy can also be written as a set of PDEs by introducing $n$ dependent fields
$\Omega^{[i]}, (i=1\dots n)$ in the following way
\begin{eqnarray}
&&U_Y=J\Omega^{[1]}\nonumber\\&&J\Omega^{[i+1]}=K\Omega^{[i]},\quad i=1\dots n-1,\nonumber\\ \quad &&U_T=K\Omega^{[n]}, \label{3}
\end{eqnarray}
and by introducing two new fields, $P$ and $\Delta$, related to $U$ as:
\begin{equation}
U=P^2,\quad\quad P_T=\Delta_X, \label{4}
\end{equation}
we can write the hierarchy in the form of the following set of equations
\begin{eqnarray} &&P_Y=-\frac{1}{2}\left(P\Omega^{[1]}\right)_X,\nonumber\\
&&\Omega^{[i]}_{XXX}-\Omega^{[i]}_X=-P\left(P\Omega^{[i+1]}\right)_X,\quad i=1\dots n-1, \nonumber \\&& P_T=\frac{\Omega^{[n]}_{XXX}-\Omega^{[n]}_X}{2P}=\Delta_X.\label{5}
\end{eqnarray}
It was shown in \cite{estevez05-1} that (2.1) can be reduced to the the negative Camassa-Holm hierarchy under the reduction $\frac{\partial}{\partial t}=0$. The positive flow can be obtained under the reduction $\frac{\partial}{\partial x}=\frac{\partial}{\partial y}$.
The conservative form of the first two equations allows us to define the following exact derivative
\begin{equation}
dz_0= P\,dX-\frac{1}{2}P\Omega^{[1]}\,dY+\Delta\,dT. \label{6}
\end{equation}
A reciprocal transformation \cite{h00}, \cite{rogers4}, \cite{rogers5} can be introduced by considering the former independent variable X as a field
depending on  $z_0$, $z_1=Y$ and $z_{n+1}=T$. From (\ref{6}) we have
\begin{eqnarray}
dX&=& \frac{1}{P}\,dz_0+\frac{\Omega^{[1]}}{2}\,dz_1-\frac{\Delta}{P}\,dz_{n+1},\nonumber\\Y&=&z_1,\quad\quad T=z_{n+1},\label{7}
\end{eqnarray}
and therefore
\begin{eqnarray}
&&X_0=\frac{1}{P},\nonumber\\&&X_1=\frac{\Omega^{[1]}}{2},\nonumber\label{8}\\&&X_{n+1}=-\frac{\Delta}{P},
\end{eqnarray}
where $X_i=\frac{\partial X}{\partial z_i}$. We can now extend the transformation by introducing a new
independent variable $z_i$ for each field $\Omega^{[i]}$ by generalizing (\ref{8}) as
\begin{equation} X_i=\frac{\Omega^{[i]}}{2},\quad i=1\dots n.\label{9}\end{equation}
Therefore, the new field $ X=X(z_0,z_1,\dots z_n,z_{n+1})$ depends on $n+2$ independent variables, where each  of the former dependent fields $\Omega_i,\,(i=1\dots n)$ allows us to define a new dependent variable $z_i$ through  definition (\ref{9}).
It requires some calculation (see \cite{estevez05-1} for  details) but it can be proved that the reciprocal transformation (\ref{7})-(\ref{9}) transforms (\ref{5}) to the following set of $n$ PDEs:
\begin{equation}-\left(\frac{X_{i+1}}{X_0}\right)_0=\left(\left[\frac{X_{0,0}}{X_0}+X_0\right]_0-\frac{1}{2}\left[\frac{X_{0,0}}{X_0}+X_0\right]^2\right)_i,\quad i=1\dots n. \label{10}\end{equation}
Note that each equation depends on only three  variables $z_0, z_i, z_{i+1}$. This result generalizes the one found in \cite{h00} for the first component of the hierarchy.
The conservative form of (\ref{10}) allows us to define a field $M(z_0,z_1,\dots z_{n+1})$ such that
\begin{eqnarray}M_i&=&-\frac{1}{4}\left(\frac{X_{i+1}}{X_0}\right),\quad\quad i=1\dots n,\nonumber\\ M_0&=&\frac{1}{4}\left(\left[\frac{X_{0,0}}{X_0}+X_0\right]_0-\frac{1}{2}\left[\frac{X_{0,0}}{X_0}+X_0\right]^2\right).\label{11}\end{eqnarray}
It is easy to prove that each $M_i$ should satisfy the following CBS  equation \cite{cbs}, \cite{cal}:
\begin{equation}M_{0,i+1}+M_{0,0,0,i}+4M_iM_{0,0}+8M_0M_{0,i}=0,\quad i=1\dots n. \label{12}\end{equation}
\textbf{The CBS equation has the Painlev\'e property} \cite {pick} and the SMM can be successfully used to derive its Lax pair \cite{estevez04}. In \cite{estevez05-1} it was proved that the Lax pair of CBS yields the following spectral problem for the CH(2+1) hierarchy (\ref{3})
\begin{eqnarray}\label{13}&&\Phi_{XX}+\frac{1}{4}\left(\lambda U-1\right)\Phi=0,\nonumber\\&&\Phi
_T-\lambda^n\Phi_Y-\frac{\lambda}{2}C\Phi_X+\frac{\lambda}{4}C_X\Phi=0.\end{eqnarray}
where $$C=\sum_{i=1}^n\lambda^{n-i}\Omega^{[i]}$$ and $\lambda(Y,T)$ is a non-isospectral parameter that satisfies
\begin{equation}\lambda_X=0,\quad \lambda_T-\lambda^n\lambda_Y=0.\label{14}\end{equation}
Consequently the problems that we meet when we try to apply the Painlev\'e test to CH(2+1) \cite{gp95} can be solved owing to  the existence of a reciprocal transformation that transforms the CH(2+1) hierarchy to $n$ copies of the CBS equation, for which the Painlev\'e methods are applicable.

\subsection{mCH(2+1) versus mCBS}
In \cite{estevez51}, one of us  introduced the following $2+1$ hierarchy (mCH(2+1) in what follows)
\begin{equation}
u_t=r^{-n}u_y, \label{15}
\end{equation}
where $r$ is the recursion operator, defined as:
\begin{equation}
r=jk^{-1},\quad k=\partial_{xxx}-\partial_x,\quad j=-\partial_x\,u\,(\partial_x)^{-1}\,u\,\partial_x.
\label{16}
\end{equation}
where $\partial_x=\frac{\partial}{\partial x}$.
 This hierarchy generalizes the one introduced by Qiao in \cite{qiao2007},
whose second positive member was studied in \cite{qiao}.
We shall  briefly summarize the results of \cite{estevez51} when a  procedure similar to that described above for CH(2+1) is applied to mCH(2+1).

If we introduce $2n$ auxiliary fields $v^{[i]}$, $\omega^{[i]}$ defined through
\begin{eqnarray}
&& u_y=jv^{[1]},\nonumber\\ &&jv^{[i+1]}=kv^{[i]},\quad \omega_x^{[i]}=uv_x^{[i]},\quad i=1\dots n-1,\nonumber\\ \quad && u_t=kv^{[n]}, \label{17}
\end{eqnarray}
the hierarchy can be written as the system:
\begin{eqnarray}
 &&u_y=-\left(u\omega^{[1]}\right)_x,\nonumber\\&&
v^{[i]}_{xxx}-v^{[i]}_x=-\left(u\omega^{[i+1]}\right)_x,\quad i=1\dots n-1,\nonumber \label{18}\\&&u_t={v^{[n]}_{xxx}-v^{[n]}_x}=\delta_x,
\end{eqnarray}
which allows to define the exact derivative
\begin{equation}
dz_0= u\,dx-u\omega^{[1]}\,dy+\delta\,dt \label{19}
\end{equation}
and $z_1=y, z_{n+1}=t$.
We can define a reciprocal transformation such that  the former independent variable $x$ is a new field $x=x(z_0,z_1,\dots \dots z_{n+1})$ depending on $n+2$ variables in the form
\begin{eqnarray}
&&x_0=\frac{1}{u},\nonumber\\&&x_i=\omega^{[i]},\label{20}\nonumber\\&&x_{n+1}=-\frac{\delta}{u}.
\end{eqnarray}
The transformation of the equations (\ref{18}) yields the system of equations
\begin{equation}\left(\frac{x_{i+1}}{x_0}+\frac{x_{i,0,0}}{x_0}\right)_0=\left(\frac{x_0^2}{2}\right)_i,\quad i=1\dots n. \label{21}\end{equation}
Note that each equation depends on only three variables: $z_0, z_i, z_{i+1}$.

The conservative form of (\ref{21}) allows us to define a field $m=m(z_0,z_1,\dots z_{n+1})$ such that
\begin{equation} m_0=\frac{x_0^2}{2},\quad m_i=\frac{x_{i+1}}{x_0}+\frac{x_{i,0,0}}{x_0},\quad i=1\dots n.\label{22}\end{equation}
Equation (\ref{21}) has been extensively studied from the point of view of  Painlev\'e analysis \cite{estevez04} and it can be considered as the modified version of the CBS equation (\ref{12}). Actually, in \cite{estevez04} it was proved that the Miura transformation that relates (\ref{12}) and (\ref{22}) is:
\begin{equation}4M=x_{0}-m \label{23}\end{equation}
A non-isospectral Lax pair  was  obtained for (\ref{22}) in \cite{estevez04}. By inverting this Lax pair through the reciprocal transformation (\ref{20}) the following spectral problem was obtained for mCH(2+1). This Lax pair reads \cite{estevez04}:

\begin{equation} \left(
\begin{array}{c} \phi \\ \hat\phi
\end{array}
\right)_x =\frac{1}{2}\left( \begin{array}{cc} -1& I\sqrt{\lambda}u \\ I\sqrt{\lambda}u
& 1
\end{array}
\right) \left(
\begin{array}{c} \phi \\ \hat\phi
\end{array}
\right),\nonumber\label{24}\end{equation}
\begin{eqnarray} \left(
\begin{array}{c} \phi \\ \hat\phi
\end{array}
\right)_t &=&\lambda^n \left(
\begin{array}{c} \phi \\ \hat\phi
\end{array}
\right)_y+\lambda a \left(
\begin{array}{c} \phi \\ \hat\phi
\end{array}
\right)_x+\nonumber \\&+&I\frac{\sqrt{\lambda}}{2}\left( \begin{array}{cc} 0& b _{xx}-b_x  \\
b_{xx}+b_x &0
\end{array}
\right) \left(
\begin{array}{c} \phi \\ \hat\phi
\end{array}
\right).\end{eqnarray}
where $$a=\sum_{i=1}^n\lambda^{n-i}\omega^{[i]}, \quad b=\sum_{i=1}^n\lambda^{n-i}v^{[i]},\quad I=\sqrt{-1}$$
and $\lambda(y,t)$ is a non-isospectral parameter that satisfies
\begin{equation}\lambda_x=0,\quad \lambda_t-\lambda^n\lambda_y=0.\label{25}\end{equation}
Although the Painlev\'e test cannot be applied to mCH(2+1), reciprocal transformations are a tool that can be used to write the hierarchy as a set of mCBS equation to which  the Painlev\'e analysis (the SMM in particular) can be successfully applied.
\section{Reciprocal-Miura transformations between CH(2+1) and mCH(2+1)}
As  stated in the previous section, there are two reciprocal transformations (\ref{8}) and  (\ref{20}) that relate CH(2+1)  and mCH(2+1)  hierarchies with CBS (\ref{12}) and mCBS (\ref{22}) respectively. Furthermore, it is known that a Miura transformation (\ref{23}) relates CBS and mCBS. The natural question that arises is whether the mCH(2+1)  hierarchy can be considered as the modified version of CH(2+1). Evidently the relationship between both hierarchies cannot be a simple Miura transformation because they are written in different variables $(X,Y,T)$ and $(x,y,t)$. The answer is provided by the relationship of both sets of variables with  the same set  $z_0,z_1,z_{n+1}$. By combining (6) and (19) we have
\begin{eqnarray}&& P\,dX-\frac{1}{2}P\Omega^{[1]}\,dY+\Delta\,dT=u\,dx-u\omega^{[1]}\,dy+\delta\,dt,\nonumber\\ &&Y=y,\quad\quad T=t, \label{26}\end{eqnarray}
which yields the required relationship between the independent variables of CH(2+1) and those of mCH(2+1). The Miura transformation (\ref{23}), also affords the following results
\begin{eqnarray} 4M_0&=&x_{0,0}-m_0\Longrightarrow \frac{X_{0,0}}{X_0}+X_0=x_0, \nonumber\\ 4M_i&=&x_{0,i}-m_i\Longrightarrow -\frac{X_{i+1}}{X_0}=x_{0,i}-\frac{x_{0,0,i}}{x_0}-\frac{x_{i+1}}{x_0}, i=1\dots n, \label{27}\end{eqnarray}
where (\ref{11}) and (\ref{22}) have been used.
With the aid of (\ref{8}), (\ref{9}) and (\ref{20}), the following results arise from (\ref{27}) (see  appendix)
\begin{eqnarray} &&\frac{1}{u}=\left(\frac{1}{P}\right)_X+\frac{1}{P},\nonumber\\ &&P\Omega^{i+1}= 2(v^{[i]}-v^{[i]}_x),\nonumber\Longrightarrow \omega^{[i+1]}=\frac{\Omega^{[i+1]}_X+\Omega^{[i+1]}}{2},\quad  i=1\dots n-1\\ &&\Delta =v^{[n]}_x-v^{[n]}.\label{28}
\end {eqnarray}
Furthermore, (\ref{26}) can be written as:
\begin{equation} dx=\left[1-\frac{P_X}{P}\right]dX+\left[\omega^{[1]}-\frac{\Omega^{[1]}}{2}\left(1-\frac{P_X}{P}\right)\right]dY+\frac{\Delta-\delta}{u}dT.\label{29}\end{equation}
The cross derivatives of (\ref{29}) imply (see  appendix) that:
\begin{eqnarray} &&\left[1-\frac{P_X}{P}\right]_Y=\left[\omega^{[1]}-\frac{\Omega^{[1]}}{2}\left(1-\frac{P_X}{P}\right)\right]_X\Longrightarrow \omega^{[1]}=\frac{\Omega^{[1]}_X+\Omega^{[1]}}{2},\nonumber\\ &&\left[1-\frac{P_X}{P}\right]_T=\left[\frac{\Delta-\delta}{P}\left(1-\frac{P_X}{P}\right)\right]_X,\Longrightarrow \frac{\delta}{u}=\left(\frac{\Delta}{P}\right)_X+\frac{\Delta}{P} \label{30}\end{eqnarray}
and therefore with the aid of (\ref{30}), (\ref{29}) reads
\begin{equation} dx=\left[1-\frac{P_X}{P}\right]dX-\frac{P_Y}{P}dY-\frac{P_T}{P}dT.\label{31}\end{equation}
This exact derivative can be integrated as
\begin{equation}x=X-\ln P.\end{equation}
By summarizing the above conclusions, we have proved that the mCH(2+1)  hierarchy
$$ u_t=r^{-n}u_y,\quad u=u(x,y,t),$$can be considered as the modified version of the Camassa- Holm hierarchy
 $$ U_T=R^{-n}U_Y,\quad U=U(X,Y,T).$$ The transformation that connects the two hierarchies  involves the reciprocal transformation
\begin{equation}x=X-\frac{1}{2}\ln U\end{equation}
as well as the following  transformation between the fields
\begin{eqnarray}&&\frac{1}{u}=\frac{1}{\sqrt U}\left(1-\frac{U_X}{2U}\right) ,\nonumber\\&&\Downarrow\nonumber \\&&\omega^{[i]}=\frac{\Omega^{[i]}_X+\Omega^{[i]}}{2},\quad i=1\dots n,\nonumber\\&& \frac{\delta}{u}=\left(\frac{\Delta}{\sqrt U}\right)_X+\frac{\Delta}{\sqrt U}\label{34}.\end{eqnarray}
\subsection{Particular case 1: The Qiao equation}
We are now restricted to the first component of the hierarchies $n=1$ in the case in which the field $u$ is independent of $y$ and $U$ is independent of Y.
\begin{itemize}
\item From (4) and (5), for the restriction of CH(2+1) we have
\begin{eqnarray}&&U=P^2,\nonumber \\&&U_T=\Omega^{[1]}_{XXX}-\Omega^{[1]}_{X},\nonumber\\&& (P\Omega^{[1]})_X=0.\label{39}\end{eqnarray}
which can be summarized as
\begin{eqnarray}\nonumber&&\Omega^{[1]}=\frac{k_1}{P}=\frac{k_1}{\sqrt U},\\&&U_T=k_1\left[\left(\frac{1}{\sqrt U}\right)_{XXX}-\left(\frac{1}{\sqrt U}\right)_{X}\right]\label{40}\end{eqnarray}
which is the Dym equation

\item The reduction of mCH(2+1)  can be achieved from  (17) and (18) in the form
\begin{eqnarray}&&\nonumber\omega^{[1]}_x=uv^{[1]}_x,\\&&u_t=v^{[1]}_{xxx}-v^{[1]}_{x},\nonumber\\&& \left(u\omega^{[1]}\right)_x=0,\end{eqnarray}
which can be written as
\begin{eqnarray}\nonumber&&\omega^{[1]}=\frac{k_2}{u},\Longrightarrow
v^{[1]}=\frac{k_2}{2u^2}\\
&&u_t=k_2\left[\left(\frac{1}{2u^2}\right)_{xx}-\left(\frac{1}{2u^2}\right)\right]_{x}\end{eqnarray}
which is the Qiao equation

\item From (29) and (33) it is easy to see that $k_1=2k_2$. By setting $k_2=1$, we can conclude that the Qiao equation.
$$u_t=\left(\frac{1}{2u^2}\right)_{xxx}-\left(\frac{1}{2u^2}\right)_{x}$$
is the modified version of the Dym equation
$$U_T=\left(\frac{2}{\sqrt U}\right)_{XXX}-\left(\frac{2}{\sqrt U}\right)_{X}$$
\item From (8) and (20), it is easy to see that the independence from $y$ implies that $X_1=X_0$ and $x_1=x_0$, which means that  the CBS and modified CBS equations (12) and (22) reduce to the following potential versions of the Korteweg de Vries and modified Korteweg de Vries equations
\begin{eqnarray}\nonumber &&\left(M_{2}+M_{0,0,0}+6M_0^2\right)_0=0,\\
&&x_2+x_{0,0,0}-\frac{1}{2}x_0^3=0.\nonumber\end{eqnarray}

\end{itemize}
\subsection{Particular case 2: The Camassa-Holm equation}
If we are restricted to the $n=1$ component when $T=X$ and $t=x$, the following results hold
\begin{itemize}
\item From (4) and (5), for the restriction of CH(2+1) we have
\begin{eqnarray}&& \Delta=P=\sqrt U,\nonumber\\ && U=\Omega^{[1]}_{XX}-\Omega^{[1]},\nonumber\\ &&U_Y+U\Omega^{[1]}_X+\frac{1}{2}\Omega^{[1]} U_X=0\end{eqnarray}
which is the Camassa Holm equation equation.

\item The reduction of mCH(2+1)  can be obtained from From (17) and (18) in the form
\begin{eqnarray}&&\delta=u=v^{[1]}_{xx}-v^{[1]},\nonumber\\&& u_y+(u\omega^{[1]})_x=0,\nonumber\\&& \omega^{[1]}_x-uv^{[1]}_x=0,\end{eqnarray}
which can be considered as a modified Camassa-Holm equation.

\item From (8) and (20), it is easy to see that $X_2=x_2=-1$. Therefore, the reductions of (12) and (21) are:
$$M_{0,0,0,1}+4M_1M_{0,0}+8M_0M_{0,1}=0,$$
which is the AKNS equation
and
$$\left(\frac{x_{1,0,0}-1}{x_0}\right)_0=\left(\frac{x_0^2}{2}\right)_1,$$
which is the modified AKNS equation.

\end{itemize}

\section{Conclusions}
In the first part of this paper previous results concerning the CH(2+1) hierarchy for a field $U(X,Y,T)$ and the mCH(2+1)  hierarchy for $u(x,y,t)$ is discussed. Reciprocal transformations that connect both hierarchies with the CBS and mCBS are constructed. The main advantage of this method is that these reciprocal transformations allow us to transform the hierarchies into a set of equations that can be studied through  Painlev\'e analysis and the methods derived from it. In particular, the Lax pairs of both hierarchies can be obtained in this way.

The Miura transformation that connects CBS and modified CBS is the key that allows us to establish a reciprocal transformation that relates the two fields $U$ and $u$ as well as the two sets of variables $(x,y,t)$ and $(X,Y,T)$. This reciprocal transformation is carefully constructed and allows us to prove that mCH(2+1)  is a modified version of CH(2+1).

As a particular case, the relationship between the two components of the hierarchies when they are independent of $Y$ and $y$ respectively is shown. It shows that the Qiao equation is a modified Dym equation.

In a similar way, we determine the modified Camassa Holm equation when the reduction $T=X$ is applied to the first component ($n=1$) of the CH(2+1) hierarchy.

\section*{Acknowledgements}
This research has been supported in part by the DGICYT under project  FIS2009-07880. We thank Professor J.M.
Cerver\'o for some interesting suggestions and a careful reading of the manuscript. We also thank the referees for their interesting suggestions and references.

\section*{Appendix}
\begin{itemize}
\item {Method for obtaining equation (\ref{29})}

Equation (\ref{27}) provides
$$x_0=X_0+\partial_0(\ln X_0).$$
If we use the fact that $X_0=\frac{1}{P}$ and $x_0=\frac{1}{u}$ we get obtain
$$\frac{1}{u}=\frac{1}{P}-\partial_0(\ln P),$$
and by using  (\ref{6}) we have
$$\frac{1}{u}=\frac{1}{P}-\frac{1}{P}(\ln P)_X$$
which yields (\ref{29}).
\item {Method for obtaining  equation (\ref{30})}

By taking $i=1\dots n-1$ in (28), we have
$$-\frac{X_{i+1}}{X_0}=x_{0,i}-\frac{x_{0,0,i}}{x_0}-\frac{x_{i+1}}{x_0}, \quad i=1\dots n-1$$
If we use (8), (9) and (20) the result is
$$-\frac{P\Omega^{[i+1]}}{2}=\partial_0(\omega^{[i]})-u\partial_{00}(\omega^{[i]})-u\omega^{[i+1]}, \quad i=1\dots n-1$$
And now (\ref{19}) gives us
$$-\frac{P\Omega^{[i+1]}}{2}=\frac{\omega^{[i]}_x}{u}-\left(\frac{\omega^{[i]}_x}{u}\right)_x-u\omega^{[i+1]}, \quad i=1\dots n-1$$
If we use the following expressions arising from (17) and (18)
$$\omega^{[i]}_x=uv^{[i]}_x,\quad u\omega^{[i+1]}=v^{[i]}-v^{[i]}_{xx},\quad i=1\dots n-1,$$
the result is
$$-\frac{P\Omega^{[i+1]}}{2}=v^{[i]}_x-v_i,\quad i=1\dots n-1$$
as  is required in (30),
and $u\omega^{[i+1]}=v^{[i]}-v^{[i]}_{xx},\quad  i=1\dots n-1$
can be written as
$$u\omega^{[i+1]}=\left(v^{[i]}-v^{[i]}_x\right)+\left(v^{[i]}_x-v^{[i]}_{xx}\right)=\left(\frac{P\Omega^{[i+1]}}{2}\right)+\left(\frac{P\Omega^{[i+1]}}{2}\right)_x,\quad i=1\dots n-1$$
and from (26), we have $\partial_x=\frac{u}{P}\partial_X$. Therefore,
$$u\omega^{[i+1]}=\left(\frac{P\Omega^{[i+1]}}{2}\right)+\frac{u}{P}\left(\frac{P\Omega^{[i+1]}}{2}\right)_X, \quad i=1\dots n-1$$
$$\omega^{[i+1]}=\left(\frac{P\Omega^{[i+1]}}{2u}\right)+\frac{1}{2P}\left(P_X\Omega^{[i+1]}+P\Omega^{[i+1]}_X\right), \quad i=1\dots n-1$$
We can eliminate $u$ with the aid of (29). The result is
$$\omega^{[i+1]}=\frac{\Omega^{[i+1]}_X+\Omega^{[i+1]}}{2}, \quad i=1\dots n-1.$$
\item {Method for obtaining equation (\ref{31})}

By taking $i=n$ in (28) we have
$$-\frac{X_{n+1}}{X_0}=x_{0,n}-\frac{x_{0,0,n}}{x_0}-\frac{x_{n+1}}{x_0}.$$
Equations (8) and (20) allow us to write the above equation as
$$\Delta=\partial_0(\omega^{[n]})-u\partial_{00}(\omega^{[n]})+\delta$$

With the aid of (19), it reads
$$\Delta=\frac{\omega^{[n]}_x}{u}-\left(\frac{\omega^{[n]}_x}{u}\right)_x+\delta$$
If $\omega^{[n]}_x=uv^{[n]}_x$ and $\delta=v^{[n]}_{xx}-v^{[n]}$ are used,
we obtain (\ref{30}). \end{itemize}

\end{document}